\documentclass[aps,prl,superscriptaddress,showpacs,floatfix,twocolumn]{revtex4}
\usepackage{times}
\usepackage{graphicx}
\usepackage{amsmath}
\usepackage{amssymb}
\usepackage{subfigure}
\usepackage{color}
\usepackage{ulem}
\usepackage{sidecap}
\usepackage{floatrow}

\begin{document}

\title{Local charge variation enhanced spin canting of BiFeO$_3$/La$_{0.7}$Sr$_{0.3}$MnO$_3$ heterostructure}

\author{Yuan-Yen Tai}
\affiliation{Theoretical Division, Los Alamos National Laboratory, Los Alamos, New Mexico 87545, USA}

\author{Jian-Xin Zhu}
\affiliation{Theoretical Division, Los Alamos National Laboratory, Los Alamos, New Mexico 87545, USA}
\affiliation{Center for Integrated Nanotechnologies, Los Alamos National Laboratory, Los Alamos, New Mexico 87545, USA}

\date{\today}

\begin{abstract}
Transition-metal oxides (TMOs) exhibit many emergent phenomena ranging from high-temperature superconductivity and giant magnetoresistance to magnetism and ferroelectricity. When TMOs are interfaced with each other, new multi-functionalities can arise, which are absent in individual components. In this work, we have systematically studied, within a unified double-exchange model,  the interfacial magnetic response in layered BiFeO$_3$ (BFO) and La$_{0.7}$Sr$_{0.3}$MnO$_3$ (LSMO) heterostructures. 
The ferromagnetic/antiferromagnetic canting is shown to be enhanced on the interface of BFO by the influence of local charge variation. More interestingly, it is found that   the spin canting in BFO can be further enhanced with deeper penetration depth to the bulk when a local oxygen vacancy is placed around the interface.
\end{abstract}

\maketitle

\paragraph{Introduction.--} Recent research on multiferroics has created lots of excitement of experimental breakthrough.
New physics such as the magnetoelectric (ME) coupling is one of the frontier research topic in the community~\cite{Cheong,Eerenstein,SMWu}.
It shows promising feature for the use of future novel device for data storage, spintronics and high-frequency magnetic devices.
However, such a multifunctional  property is difficult to realize under a single phase crystal~\cite{Prellier}.
In recent years, although a handful of multiferroic systems have been experimentally realized,  quite often these multiferroics come
either with a very low Curie temperature ($T_c$) for ferromagnetic order~\cite{santos}
or with a high Neel temperature ($T_N$) for antiferromagnetic order~\cite{teague}.
So far, the transition-metal oxides (TMOs) BiFeO$_3$ (BFO)~\cite{vorob,zhao,jang,wang}
and $R$MnO$_3$ (where $R$ for rare-earth elements) are candidates for multiferroics~\cite{kimura_nature,kimura_PRB}.
However, none of these single-phase perovskite materials demonstrate significant and robust electric and magnetic polarizations at room temperature. In particular, materials like BFO and TbMnO$_3$ exhibit either commensurate or sinusoidal antiferromagnetism, a reason why BFO is sometimes called a ferroelectric antiferromagnetic material. This undesired property limits their potential technological applications. Interfacial engineering of  complex oxide nanostructures with controlled geometry and dimensionality can provide an unprecedented platform to induce new electronic and magnetic properties.  Synthesis of high quality TMO heterogeneous structures,  and characterization of their magnetic$/$electric properties are the current focus of experimental activities~\cite{Niebieskikwiat,YMKim}.  
The study of  multiferroic BFO  and ferromagnetic La$_{0.7}$Sr$_{0.3}$MnO$_3$ (LSMO) interface has shown the improvement of magnetic property, which is crucial to technological applications~\cite{PYu,SSingh}. The induced ferromagnetism in the BFO side of BFO/LSMO interfaces has also been demonstrated theoretically within density functional theory~\cite{SSingh,RNeumann} and a tight-binding model analysis~\cite{Calderon}. The \textit{ab-initio} study~\cite{SSingh} has also shown that the magnetic coupling between BFO and LSMO is sensitive to the nature of the atomic layer at the interface. Also very interestingly, it has been  observed~\cite{YMKim} that the accumulation of vacancies could strongly affect the polarization switching phenomena, suggesting that the microstructure of the heterogeneous compound play important roles for the ME coupling~\cite{AChen,Krishnan}.
In this Letter, we propose a generalized double-exchange model to uncover the mechanism of the spin canting as well as its penetration depth into the BFO when it is formed with LSMO into a planar heterostructure. Particularly, we will consider the role of Coulomb interaction from both the itinerant electrons and ionic atoms and that of the microstructure by introducing oxygen vacancy. 

\paragraph{Model setup.--}
We consider a model within a double-exchange mechanism through the bridge of oxygens and the screened Coulomb interaction in the perovskite lattice structure.
There the t$_{2g}$ electrons are fully localized and treated as classical spins, which are coupled to the itinerant e$_{g}$ electrons via a finite Hund's coupling.
\begin{equation}
\begin{aligned}
\label{Ham}
H = & \sum_{Ib\, m\, \sigma} t^{(pd\sigma)}_{Ib} O^{1}_{m\, b}\; ( d^\dagger_{I\, m\, s} p_{Ibs} + p^\dagger_{Ibs} d_{I\, m\, s})\\
		+ & \sum_{I,b\neq c,s}t^{(pp)}_{Ibc} O^{2}_{b-c} \;p^\dagger_{Ibs} p_{Ics}\\		
        - & \sum_{J\, m\, ss'} J^H_I\, \vec S_I \cdot \vec \sigma_{ss'} d^\dagger_{I\, m\, s} d_{I\, m\, s}\\
        + & \sum_i (\phi_i + \varepsilon_i - \mu)n_i\;.
\end{aligned}
\end{equation}
Here $t^{(pd\sigma)}_{Ib}$ ($O^1_{m\, b}$) and $t^{(pp)}_{Ibc}$ ($O^2_{b-c}$) describes the Slater-Koster parameters~\cite{SK} for $dp$- and $pp$-orbital hoppings in the perovskite lattice; the index $m \in [d_{x^2-y^2}, d_{z^2-r^2}]$;
$J^H_I$ is the Hund's coupling between the local t$_{2g}$  spins, $S_J$, and itinerant e$_g$  spin density, $\vec\sigma_{ss'}$;
the index $I$ denote each site for the transition-metal (Fe$/$Mn) and $b,c\in [\pm\frac{a}{2}\hat x, \pm\frac{a}{2}\hat y, \pm\frac{a}{2}\hat z]$ indicate the surrounding oxygen sites; the last term in Eq.~(\ref{Ham}) describes the screened Coulomb potential, $\phi_i$, on-site energy, $\varepsilon_i$, for each atom (Bi$/$La(Sr)$/$Fe$/$Mn$/$O) at position $\vec r_i$, and  the chemical potential, $\mu$.
The screened Coulomb potential is set as,
\begin{equation}
	\phi_i = \sum_j  \frac{\alpha(n_j - Z_j)}{|\vec r_i - \vec r_j|}\times e^{-|\vec r_i-\vec r_j|/r_0}\;.
\end{equation}
In principle, $t^{(pd\sigma)}_{Ib}$, $t^{(pp)}_{Ibc}$ and $J^H_I$ depend on materials, and can be influenced by the lattice distortion, strain, and microstructure near the interface. We simplified the problem by assuming they are uniform: $t^{(pd\sigma)}_{Ib}=t^{(pd\sigma)}$, $t^{(pp)}_{Ibc}=t^{(pp)}$ and $J^H_I = J^H$.
Therefore, the difference for BFO and LSMO is addressed only by the on-site energy difference, $\varepsilon_i$, the ionic charge, $Z_i$, and the local electron filling, $n_i$,  which are specified in Table~\ref{table:parameters}. In this table, the electron occupancy $n_i$ (given in the single phase) of each Fe/Mn/O atoms is highly dependent on the given values of $\varepsilon_i$, and it will be varying if the translational symmetry broken, e.g., in the case of a heterostructure as considered in the present work.

\begin{table}
\begin{tabular}{c | c c  c  c  c}
	      & \;\;Bi\;\; & La(Sr) & \;\;Fe\;\; & \;\;Mn\;\; & \;\;O\;\; \\
\hline
 $\varepsilon_i$ & --- & \;---     & \;0          & \;4            & \;-10	\\
 $Z_i$      & 3     & \;2.7  & \;5       & \;4         & \;0   \\
 $n_i$      & 0     & \;0    & \;$\sim$2 & \;$\sim$0.7 & \;$\sim$2   \\
\end{tabular}
\caption{On-site energy, charge, and electron occupation (in the single phase) of different atomic specifies. Here Bi, La(Sr) are treated as pseudo atoms that their orbital degree of freedom are not involved in the Hamiltonian of Eq.~\ref{Ham}, however, these atoms act as a attractive center at $\vec r_i$ with, $n_j=0$.
Specifically, we treated La and Sr atoms as a single atom that carry a mixed fractional charge, $Z=2.7$ for the doping $x=0.3$.}
\label{table:parameters}
\end{table}

We solved Eq.~(\ref{Ham}) self-consistently for the optimized charge distribution and spin orientation for a given BFO/LSMO interface structure.
Firstly,the chemical potential, $\mu$, is adjusted at each iteration to fulfill the conservation of the total number of electrons, $\sum_i n_i = n_{total} =  \text{num(Fe)}\times 8 +\text{num(Mn)}\times 6.7$.
Secondly, the screened Coulomb potential, $\phi_i$, is determined via the  self-consistently determined local electron density. 
Finally, after the relaxation of the local $e_g$ charge density, the local spin structure is  solved by {\it Langevin-Landau-Gilbert} (LLG) spin dynamics,
\begin{equation}
\label{LLG}
\frac{d\vec S_I}{dt} = \vec S_I\times \vec F_I+\eta(\vec S_I\times \vec F_I)\times \vec S_I,
\end{equation}
where $F_I$ is the effective field from the Hellmann-Feynman theorem,
$\vec F_I = -\frac{\partial E}{\partial \vec r_I} = -\sum_{mn} \rho_{mn} \frac{\partial H_{mn}}{\partial \vec r_I}$,
where the $\rho_{mn}= f_{\text{FD}}(H)_{mn}$ is the density matrix and $f_{\text{FD}}(x)$ is the Fermi-Dirac distribution function.
In Eq.~(\ref{LLG}), $\eta$ is a positive value for the damping term and we set $dt=0.1$ to update the local spin orientation.
Throughout the work,  the following empirical values were chosen for the parameters: $t^{(pd\sigma)}=1.7$, $t^{(pp)}=0.5$, $J^H=\frac{4}{3}$; while  $\alpha$ will taken different values under different circumstances. These hopping parameters were referenced  from Ref.~\onlinecite{Mostovoy}. 
The screened Coulomb potential contributed from all the surrounding atoms up to five lattice constant was taken into account and the screened radius is set to be $r_0=2.5a$, where $a$ is the lattice constant. Hereafter, all energies are measured in units of  eV while the length is measured in units of $a$.

\paragraph{Results.--}

\begin{figure}
\includegraphics[scale=0.15]{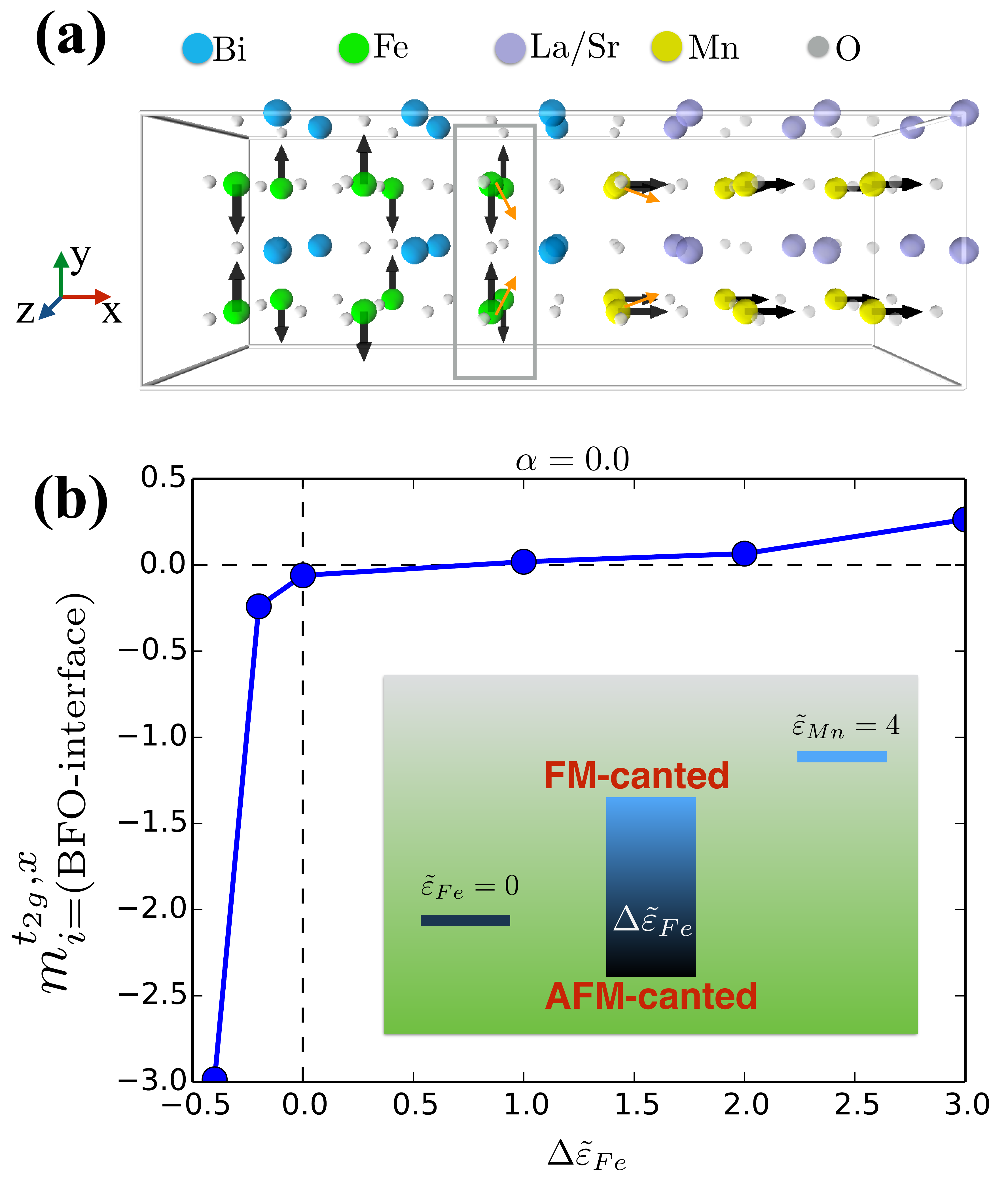}
\caption{(Color online) 
(a) A schematic drawing of an planar BFO/LSMO heterostructure,  and an initial spin configuration (black arrows) for a (3+3)$\times$2$\times$2 unit-cell system (3-layers of BFO and 3-layers of LSMO). The red arrow shows a possible spin configuration after the LLG spin dynamics relaxation. The {\bf interface-Fe} (i-Fe) atoms are indicated in the rectangle.
(b) The calculated spin polarization for a (7+7)$\times$2$\times$2 layered BFO/LSMO heterostructure as a function of an \textit{ad hoc} 
on-site energy $\Delta\tilde\varepsilon_{Fe}$ for the i-Fe $e_g$-orbital. 
The $m^{t_{2g},x}_i$  indicates the $x$-component of a fully localized t$_{2g}$-spin on site $i$. }
\label{fig01}
\end{figure}

We first examined the spin structure for  the single phase BFO$ and $LSMO, respectively,  in a 2$\times$2$\times$2 perovskite unit cell. 
When  the Hund's coupling in Eq.~\ref{Ham} is turned off, i.e., $J^H = 0$, the band structure shows a metallic state for both BFO and LSMO, respectively.
Once we turned on the Hund's coupling, an insulating-AFM for BFO while a metallic-FM phase  for LSMO is obtained
after the relaxation of the fully localized t$_{2g}$ spin with the LLG dynamics.
The band gap for BFO mostly depends on the strength of the Hund's coupling.
and for  $J^H=\frac{4}{3}$,  a band gap  of about  3.5 eV was found.
This benchmark allows us to confirm that out model setting and the implementation of LLG spin dynamics are built upon a firm base.

We now turn to  the BFO-LSMO heterostructure, where the Coulomb potential starts to play an important role in our model.
The local electron density, $n_i$, will provide a feedback to $\phi_i$, and subject a renormalized on-site potential, $\tilde\varepsilon_i \equiv \phi_i+\varepsilon_i$. To gain a qualitative understanding of this finite-ranged Coulomb interaction, we set $\alpha=0$ but instead set all the bulk $\tilde\varepsilon_{Fe}=0$, $\tilde\varepsilon_{Mn}=4$ except for the on-site energy level of the {\bf interface-Fe} (i-Fe) atoms with $\Delta\tilde\varepsilon_{Fe}$.
Fig.~\ref{fig01}(a) shows a schematic setup of a smaller (3+3)$\times$2$\times$2 unit cell (3-layers of BFO and 3-layers of LSMO).
The actual calculation in Fig.~\ref{fig01}(b) was performed for  a (7+7)$\times$2$\times$2 unit-cell heterostructure.
The initial spin structure for BFO and LSMO was set as shown in Fig.~\ref{fig01}(a), and it was then relaxed every site with the LLG spin dynamics. Finally, we can find the i-Fe atoms are FM$/$AFM canted with LSMO when the $\Delta\tilde\varepsilon_{Fe}$ is positively$/$negatively adjusted.
Fig.~\ref{fig01}(b) also shows the nonlinear behavior for the FM$/$AFM canting as afunction of $\Delta\tilde\varepsilon_{Fe}$, and it indicates the model system is easier to push the interface BFO layer to be AFM canted.

\begin{figure}
\includegraphics[scale=0.15]{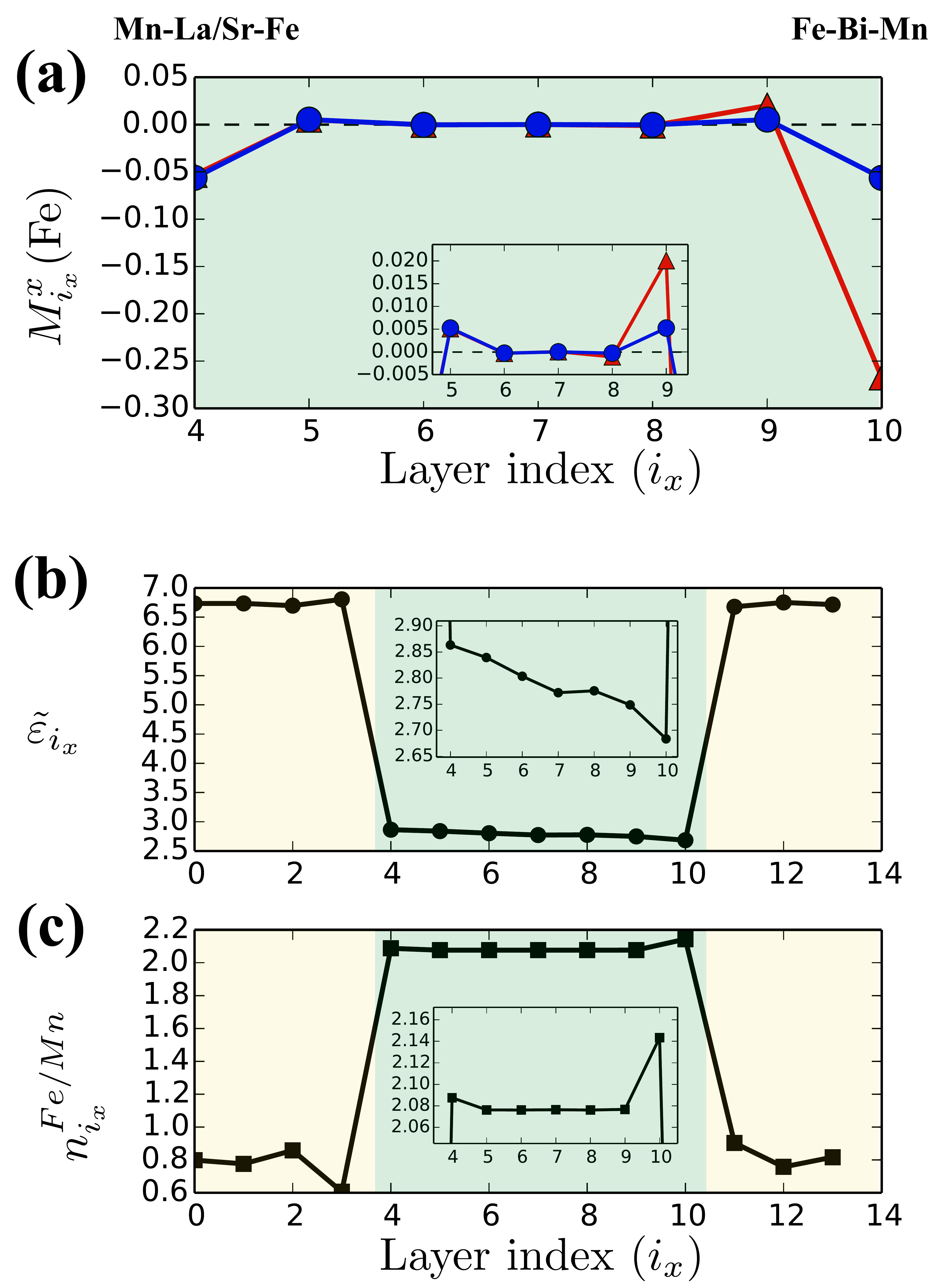}
\caption{(Color online) 
Calculated $x$-component of the total spin, $M^x_{i_x}$, on BFO-Fe atoms (for $\alpha=0$ and $0.4$) (a), renormalized $e_g$-orbital on-site energy level (b), and electron occupation $n^{Fe/Mn}_{i_x}$ (c) in a 
(7+7)$\times$2$\times$2 BFO/LSMO heterostructure. The green and yellow shaded regions indicate the BFO and LSMO layers, respectively.
 The left-hand side is interfaced through La/Sr virtual ions and the right-hand side is interfaced through Bi virtual ions. For panels (b) and (c), only the result for $\alpha=0.4$ is shown.
The inset in each panel is a  zoom-in display of the results from layer 4 through 10. 
}
\label{fig02}
\end{figure}

\begin{figure}
\includegraphics[scale=0.12]{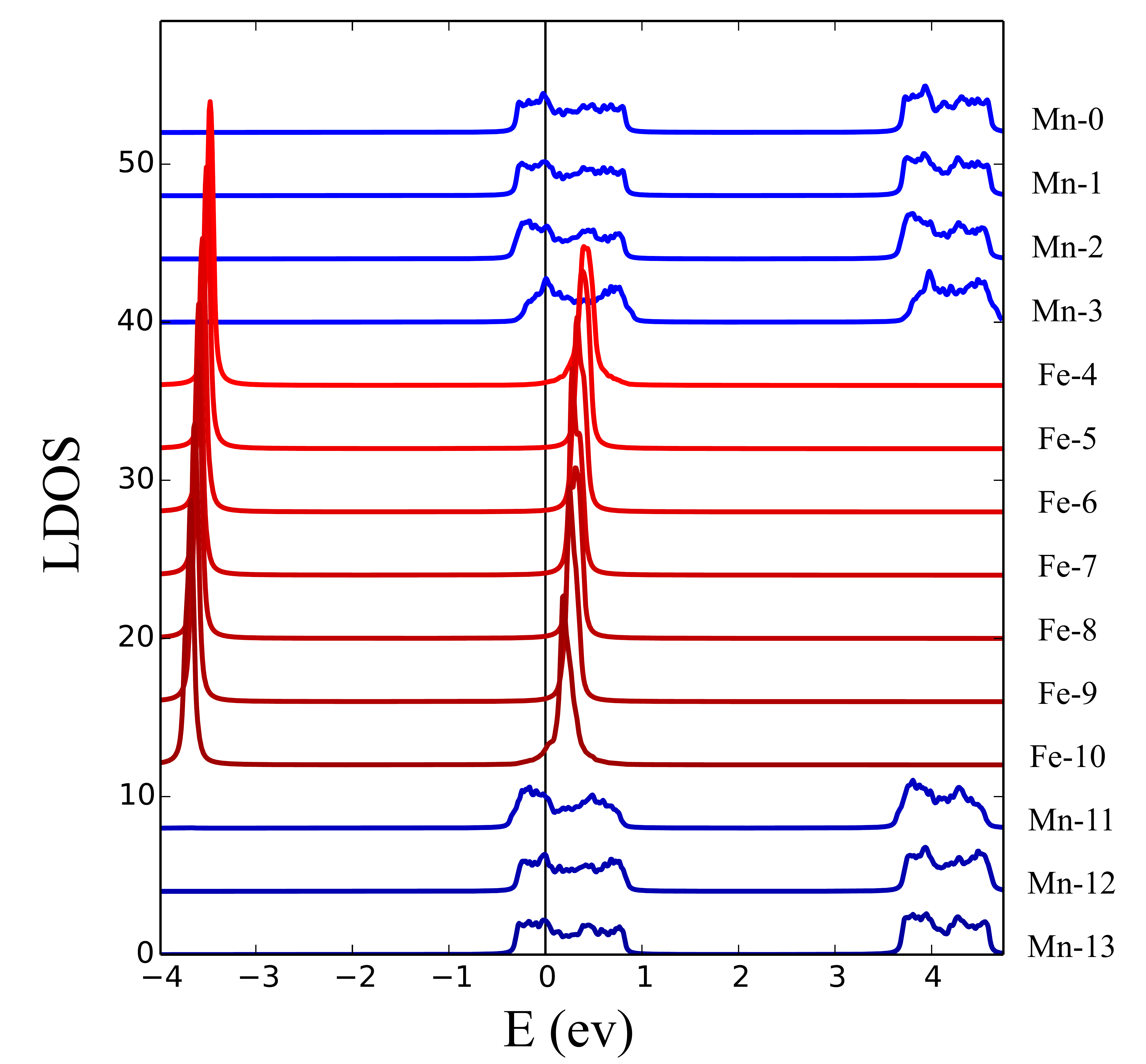}
\caption {
(Color online) Layer dependence of the Fe$/$Mn $e_g$-orbital LDOS characteristic throughout the (7+7)$\times$2$\times$2 BFO/LSMO heterostructure with $\alpha=0.4$.
}
\label{fig03}
\end{figure}

For an actual Coulomb potential  (not artificial adjustment for the local on-site energy as discussed above), the spin and charge profile is no longer symmetric along $x$-direction when $\alpha\neq 0$ with respect to the center of BFO bulk, which are placed from 4th to 10th layers in the (7+7)$\times$2$\times$2 unit-cell heterostructure (see Fig.~\ref{fig02}). 
This asymmetry arises from the fact that 
Fe and Mn atomic layers at the two interfaces sandwiches Bi and La/Sr atomic layers, which have different ionic charges, $Z$. 
Since  the translational invariance in $y$- and $z$- directions is still preserved, the spins in each layer of BFO and LSMO are uniformly canted to the same side (AFM or FM).
Figure~\ref{fig02} shows the results on the  $x$-component of measurable total magnetic polarization for each layer (averaged over the $y$- and $z$-components of the coordinates), $M^x_{i_x}\equiv \frac{1}{N_{yz}}\sum_{i_{y,z}}\Big[ m_{i}^{t_{2g},x}+m_{i}^{e_g,x} \Big]$, where $m_i^{e_g,x}\equiv \sum_{m_{l}}\langle \sigma_{im_{l}}^{e_{g},x} \rangle$ with $\sigma$ being the Pauli matrix and $m_l$  denoting the two $e_g$-orbitals, effective  $e_g$-orbital on-site energy  and local electron density in the (7+7)$\times$2$\times$2 unit-cell heterostructure.
As shown in Fig.~\ref{fig02}(a), for $\alpha=0$,  the result is symmetric along $x$ as expected.
The spins of BFO are AFM-canted with respect to those of LSMO layers across the interfaces.
The maximum canting take places on the first layer of BFO at the interfaces.
With the introduction of the Coulomb interaction (e.g., $\alpha=0.4$ in Fig.~\ref{fig02}), 
one can find out that the AFM canting is significantly enhanced in the first layer of BFO in the interface with the Bi  atoms, while remains not much changed in the interface with La(Sr) atoms. This is due to the fact that the  Bi ions are more attractive than the mixed La(Sr) ions, causing stronger  charge variation at  the interface.  The calculated renormalized $e_g$-orbital on-site energy levels and electron density provides a clear evidence for this phenomenon (see Fig.~\ref{fig02}(b)-(c)). Figure~\ref{fig03} displays the layer dependent Fe/Mn $e_g$-orbital local density of states (LDOS) in the  (7+7)$\times$2$\times$2 BFO/LSMO heterostructure  with $\alpha=0.4$. Noticeably, we see that although the single-phase BFO is an electronic insulator, the top layers of BFO at the interface of the BFO/LSMO heterostructure exhibit a metallic behavior, as evidenced by the finite LDOS intensity at the Fermi energy (see the 4th and 10th-layer of the heterostructure). 
This is related to the Coulomb potential induced charge accumulation in the BFO layers near the interfaces.
However, as the induced ferromagnetism, this metallic behavior decays rapidly into the BFO bulk, where an insulating behavior is recovered.

\begin{figure*}
\includegraphics[scale=0.15]{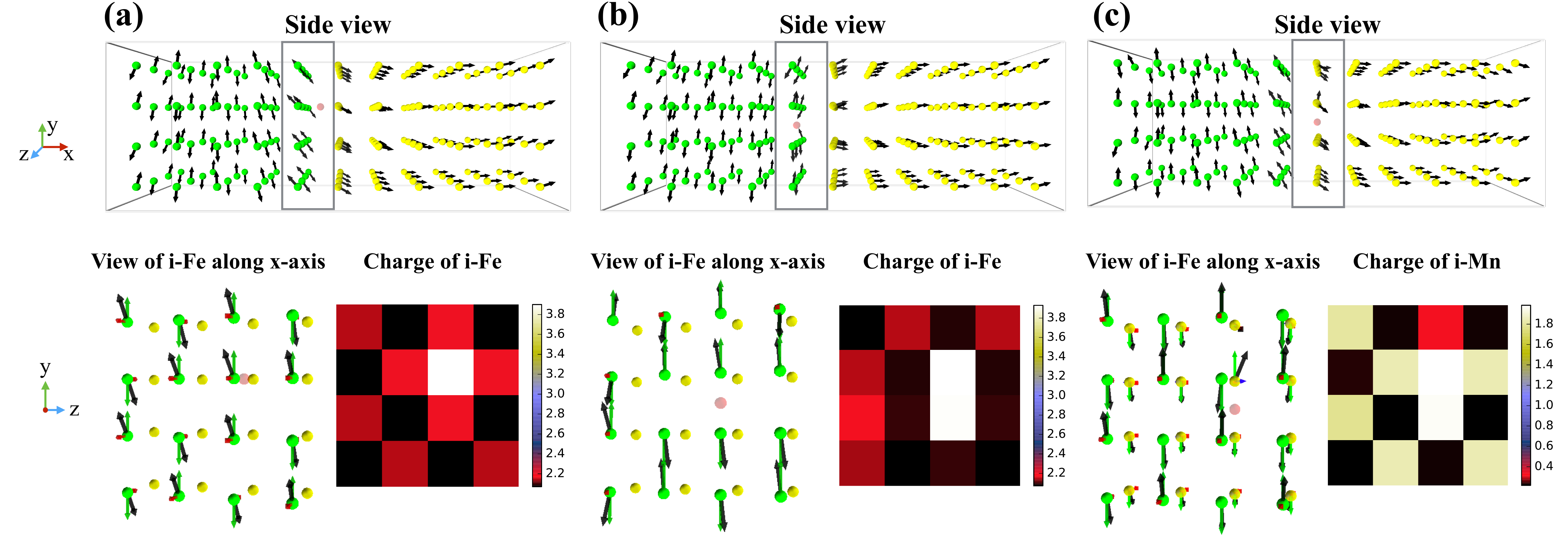}
\caption{(Color online) 
Spin and $e_g$-orbital charge profile of a BFO/LSMO heterostructure with an oxygen vacancy located in between the bond of Fe and Mn layers (a), in between the bond of the interface layer of Fe-Fe atoms (b), and in between the bond of the interface layer of Mn-Mn atoms (c). The top panels are for the side view of three-dimensional spin texture on Fe (green) and Mn (yellow) sites. The oxygen vacancy is denoted by a red ball.
The bottom panels show the view, along the $x$-direction,  of the spin (left)  and $e_g$-orbital charge (right) distribution on the interface layer. 
The red (green) arrows in the left-bottom panel of each case show  the in-plane projection of spin vectors.}
\label{fig04}
\end{figure*}

\begin{figure}
\includegraphics[scale=0.18]{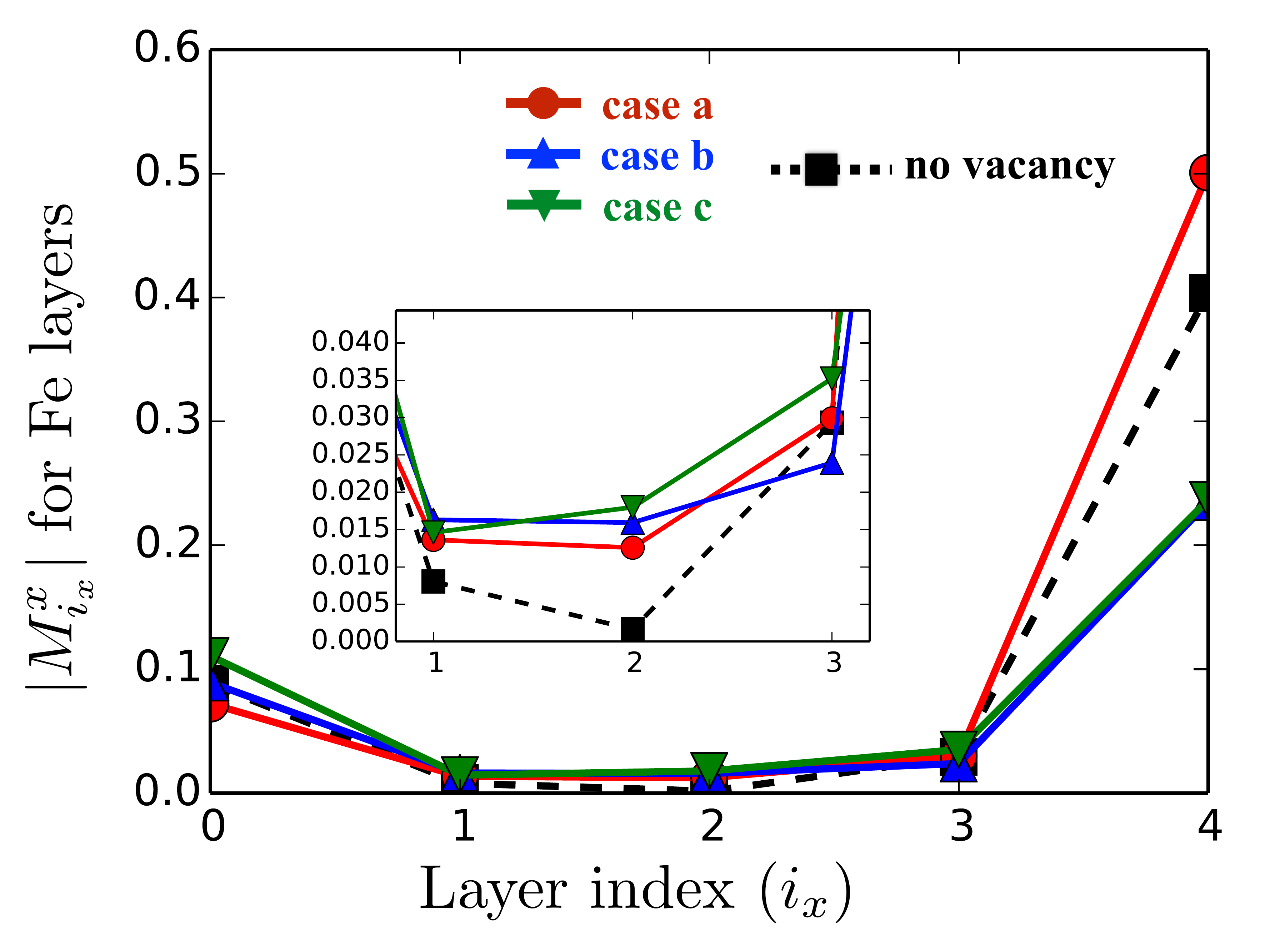}
\caption{(Color online)
Ferromagnetic moment due to the  spin canting inside the BFO layers for the three configurations of O-vacancy. 
The labels a, b and c correspond to those in Fig.~\ref{fig04}.
The black-dashed (square) line denote the result without any vacancy.
The inset is a zoom-in display of results in layers 1 to 3 of BFO.
}\label{fig05}
\end{figure}

The above situation can be changed when a microstructure like oxygen vacancy around the interface is introduced. For this purpose, we increased the linear dimension along the $y$- and $z$-direction of the supercell,  and considered a single oxygen vacancy in a  (5+6)$\times$4$\times$4 heterostructure (5 BFO and 6 LSMO layers). In the absence of O-vacancy, the Coulomb interaction from the oxygen atoms will provide  a homogeneous negative background with electron fillings close to $2$, and there is only a very small charge variation near the BFO$/$LSMO interface.
Once an  O-vacancy is introduced, the local $dp$- and $pp$-orbital hoppings will be quenched, and  more electrons will be accumulated around its surrounding partially filled $e_g$-orbitals of Fe and Mn atoms. In the following, we present the results for three spatial configurations of O-vacancy in the heterostructure. 
 When the vacancy is located in between the bond of Fe and Mn atoms (c.f. Fig.~\ref{fig04}(a)), it generates an oscillating pattern of the Fe charges ($n_{i=\text{i-Fe}}$) with a C$_4$ symmetry.
Due to this local charge redistribution, the AFM and FM canting (with respect to the LSMO magnetization direction) on each Fe atom are strongly enhanced and the average AFM canting is $|M^x_{i_x=\text{i-Fe}}| \simeq 0.5$ (See  the red circle on $i_x = 4$ in Fig.~\ref{fig05}).    Similar behavior is also obtained for O-vacancy in the BFO side of the interface (c.f. Fig.~\ref{fig04}(b)) and O-vacancy in the LSMO side of the interface (c.f. Fig.~\ref{fig04}(c)), except that  the C$_4$ symmetry of the charge distribution is no longer preserved.
For the first two O-vacancy cases, the highest charge density of the i-Fe layer is located on those atoms closest to the O-vacancy, while for the third O-vacancy configuration, the highest charge density of the i-Fe  is located in the corner rather than those closest to the O-vacancy. 
This shows that not only the local charge can influence the spin canting on each Fe site, the spin structure can also have a back action effect on the local charge.
Specifically, when the O-vacancy is located in the  LSMO side of the interface,  the charge variation  no longer follows the same scenario as (a) and (b) and instead it must occur through the  double exchange interaction.
Although the majority of the spin component of LSMO is still aligned to the $x$-direction, different location of the O-vacancy can result in different noncollinear spin structure in LSMO.
Finally, we display in Fig.~\ref{fig05} the magnetic moment distribution in the BFO layers for the three configurations of the O-vacancy.
One can see  that although the O-vacancy does not enhance too much the induced ferromagnetism on the i-Fe layer, due to the cancellation of AFM and FM canting of each Fe site (with respect to the LSMO ferromagnetization),  
the presence of O-vacancy can increase the penetration of the ferromagnetism into the bulk.
This result can point to a right direction to explain for the experimentally observed finite ferromagnetism even deep into the BFO bulk in a BFO/LSMO superlattice~\cite{SSingh}.

\paragraph{Concluding Remarks.--}
In summary, we have used a generalized double exchange model to study the electronic and magnetic properties of BFO/LSMO heterostructures. We have elucidated the importance of Coulomb interaction in explaining interesting magnetic phenomena at the interface. Furthermore, our generalized model enables a direct study of the microstructure effect on the magnetic properties of TMO heterostructures. We have considered a oxygen vacancy near the interface and shown clearly that the it could modify the range of the spin canting into the BFO bulk.  Our study has provided a convincing evidence that the oxygen vacancy should be considered as a new control parameter in designing the functionality of TMO nanocomposites via the charge-spin coupling.  Similarly, the doping of other atoms like F should also be able to modify the functionality. We also remark that the inclusion of the Heisenberg-like superexchange coupling is straightforward in our model. Although a weak superexchange coupling will not have a significant influence on the magnetism in the BFO side of the heterostructure,  it will further enhance the noncollinear spin behavior on the LSMO side.

\paragraph{Acknowledgements.--}
We thank Hongchul Choi, Satoru Hayami, and M. Mostovoy for useful discussions.
This work  was supported by the U.S.  DOE Contract No.~DE-AC52-06NA25396 through the
LDRD Program (Y.-Y.T.). This work was supported in part by the Center for Integrated Nanotechnologies, a DOE BES user facility (J.-X.Z.).

\end{document}